\documentclass[10pt,conference]{IEEEtran}

\IEEEoverridecommandlockouts
\usepackage{cite}
\usepackage{amsmath,amssymb,amsfonts}
\usepackage{algorithmic}
\usepackage{graphicx}
\usepackage{textcomp}
\usepackage{xcolor}
\usepackage{float}
\usepackage{booktabs}
\usepackage{multirow}
\usepackage{comment}
\usepackage{url} 
\def\BibTeX{{\rm B\kern-.05em{\sc i\kern-.025em b}\kern-.08em
    T\kern-.1667em\lower.7ex\hbox{E}\kern-.125emX}}

\begin{document}

\title{Estimating Cellular Network Delays in Finnish Railways: A Machine Learning Enhanced Approach
}

\author{
    \IEEEauthorblockN{
        Saeideh Mansouri\textsuperscript{1}, Mohamed Shamekh\textsuperscript{2}, Simon Indola\textsuperscript{3}, Petri Mähönen\textsuperscript{1}
    }
    \IEEEauthorblockA{
        \textsuperscript{1} Department of Information and Communications Engineering, Aalto University, Espoo, Finland. \\
        Emails:  \{ Saeideh.Mansouri, Petri.Mahonen \}@aalto.fi \\
        \textsuperscript{2} Omnitele , Helsinki, Finland. \\
        Email: Mohamed.Shamekh@omnitele.com \\ 
        \textsuperscript{3} Fintraffic , Helsinki, Finland. \\
        Email: simon.indola@fintraffic.fi 
    }
}

\maketitle

\begin{abstract}
There is growing interest in using public cellular networks for specialized communication applications, replacing standalone sector-specific networks. One such application is transitioning from the aging GSM-R railway network to public 4G and 5G networks. Finland is modernizing its railway communication system through the Digirail project, leveraging public cellular networks. To evaluate network performance, a nationwide measurement campaign was conducted in two modes: Best Quality and Packet Replication. However, Best Quality mode introduces artificial delays, making it unsuitable for real-world assessments. In this paper, railway network delays are modeled using machine learning based on measurements from the Packet Replication mode. The best-performing model is then employed to generate a dataset estimating network delays across Finland’s railway network. This dataset provides a more accurate representation of network performance. Machine-learning-based network performance prediction is shown to be feasible, and the results indicate that Finland’s public cellular network can meet the stringent performance requirements of railway network control.
\end{abstract}


\section{Introduction}

Mobile communication systems are essential for the successful operation of rail networks. They provide a critical link between ground control and trains, enabling reliable two-way communication. The Global System for Mobile Communications – Railway (GSM-R) has been the backbone of railway communications for years. However, as operational demands increase and GSM-R becomes obsolete, more advanced technologies such as 4G and 5G are being explored \cite{cansu2023,Mazzenga2017}. These next-generation networks offer higher data rates, lower latency, and improved reliability, which are key enablers for autonomous train operations and advanced rail network management. LTE and 5G can also support services such as high-definition video monitoring and diagnostics, extending the capabilities of the European Rail Traffic Management System (ERTMS) \cite{zhang2022,laroche2013ertms,IDATE2015}.

Recent studies emphasize the urgency of transitioning from GSM-R to modern communication frameworks such as the Future Railway Mobile Communication System (FRMCS), based on LTE and 5G, to accommodate increasing data demands and support new applications like autonomous operation and real-time monitoring \cite{analysys2014,grilli-phd}. This transition poses both technical and strategic challenges, requiring close collaboration among European railway stakeholders to ensure seamless migration and cross-border interoperability \cite{IDATE2015,grilli-phd}.

Railways incur significant financial losses by not leveraging public mobile networks for train control and management signaling \cite{EDel2014, Mazzenga2017}. In \cite{EDel2014}, the authors propose integrating Public Land Mobile Networks (PLMNs) with a satellite backup to replace GSM-R. Their approach improves availability through dynamic network selection, reduces reliance on single providers, and lowers costs compared to GSM-R. They also suggest extending the Euroradio protocol for seamless multi-bearer communication.

Mazzenga et al. \cite{Mazzenga2017} advanced the ideas from \cite{EDel2014} by implementing a practical PLMN-satellite integration using Multi-path Transmission Control Protocol (TCP). In a 10,000 km test over a 300 km railway line, they demonstrated reliable connectivity and low latency. Their intelligent routing strategy enhances Quality of Service (QoS) and enables cost-effective modernization of regional rail networks.

Carro-Lagoa et al. \cite{carro-lagoa_feasibility_2019} evaluated the feasibility of LTE for subway train control in the Madrid Metro, focusing on Communication-Based Train Control (CBTC) and video surveillance. While LTE met CBTC latency and data rate requirements, uplink delays impacted video quality. The study emphasized the need for QoS optimization, better handovers, and interference management to ensure reliable operation.

Finland is modernizing its railway communication system through the Digirail project, integrating public 4G and 5G networks to support the European Train Control System (ETCS). There are three main mobile network operators in Finland: Elisa, Telia, and DNA. All three can provide full coverage along the railway tracks. In some regions, two operators share parts of their physical network infrastructure. To evaluate whether the networks provided by these operators can meet the latency requirements of the railway communication system, a nationwide measurement campaign was carried out across Finland’s railway network. The key objective was to assess whether these separate mobile networks could independently meet the required latency bounds. This could be achieved either by identifying one or two operators capable of fulfilling the requirement alone or by using multihoming (packet replication) to combine multiple networks and ensure the necessary operational QoS.

Hence, two measurement methods were used to evaluate radio network performance: the entire railway was initially measured using the Best Quality (BQ) mode, and selected segments were subsequently remeasured using the Packet Replication (PR) mode. The main drawback of BQ mode is that it assesses network conditions only once every five seconds, transmitting data via the network with the strongest signal at that moment. However, network conditions may change before the next assessment, introducing potential delays. Additionally, the network status check itself causes a minor delay. In contrast, PR mode minimizes delays by simultaneously sending IP packets through all available radio connections, selecting the first packet to arrive as the valid data point, and discarding the rest. This approach mitigates delays caused by fluctuating network conditions \cite{indola_suitability_2023}.

Results showed that public mobile networks met ETCS requirements for reliability, latency, and jitter. In PR mode (Diversity mode), the system achieved 99.986\% reliability. While legacy Subset-093 requirements were easily met, compliance with FRMCS standards may require additional network investments. The study concluded that public mobile networks, combined with multi-channel routers, offer a cost-effective and reliable alternative for ETCS traffic delivery \cite{indola_suitability_2023}.

Since only 2,000 km of the railway network were measured in PR mode, and repeating these measurements for the remaining segments was impractical, this paper first models PR-mode delays using Machine Learning (ML) methods. Then, the best-performing model is applied, for each delay type, to generate a PR-mode dataset for the parts of the Finnish railway network that were measured in BQ mode. Moreover, datasets generated by ML methods can help reduce the need for extensive future measurement campaigns. It allows future measurements to be conducted in faster or simpler modes and enables reduced sampling, as the developed ML-based delay models have proven to be highly reliable for statistical analysis and modeling purposes.


The remaining part of the paper is organized as follows: Section II reviews the dataset and ML methods. Section III presents modeling results and compares generated and measured PR-mode data. Section IV evaluates network reliability, and Section V concludes the paper.

\section{Measurement Dataset and ML-Approach}

The Digirail radio network measurement system is designed to evaluate the transmission of simulated ETCS traffic over public mobile networks. The collected data includes Global Navigation Satellite System (GNSS) positioning, train speed, and signal quality indicators such as Reference Signal Received Power (RSRP), Reference Signal Received Quality (RSRQ), and Signal-to-Noise Ratio (SNR). It also captures network performance metrics, including one-way end-to-end delays for TCP connections, Hypertext Transfer Protocol (HTTP) requests, and Domain Name System (DNS) queries. Additionally, simulated ETCS-related data, such as Position Report and Movement Authority (MA) requests, are recorded. All parameters are sampled at 1-second intervals, except for MA requests, which are sampled every 10 seconds. The measurement system utilized a multi-channel router with multiple SIM cards to test various operators, with all measurements conducted in 4G/LTE mode\cite{indola_suitability_2023}.

\begin{figure}[htbp]
    \centering
    \includegraphics[width=0.35\textwidth]{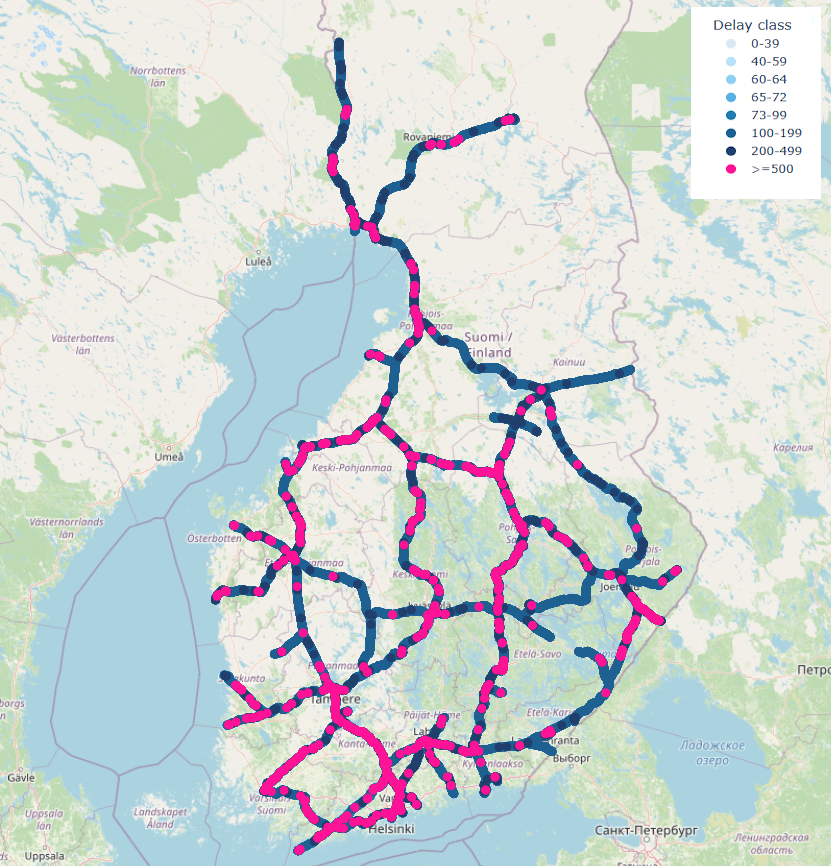}
    \caption{Position Report in Best Quality mode.}
    \label{fig:position_BQ}
\end{figure}

To illustrate the measurement coverage, Figure \ref{fig:position_BQ} shows track segments measured in BQ mode. Data points represent Position Report delays, with values below 500 milliseconds shown in light to dark blue, and delays exceeding 500 milliseconds highlighted in pink to indicate critical events. As expected, a significant number of critical points appear in the BQ-mode dataset, reflecting the limitations of this configuration.

To model PR-mode delays, four ML algorithms are evaluated, and the best-performing model for each delay type is selected to extrapolate PR-mode data across Finland’s railway network. This task is framed as a supervised regression problem, where input features comprise network KPIs from all three operators and train speed, constrained to PR-mode data. However, the primary focus is {\em regression}; classification performance is assessed by evaluating precision and recall. Specifically, model accuracy in identifying critical values (instances where predictions exceed predefined thresholds) is examined. Critical points are defined as delays exceeding 500 ms for Position Report, MA request, TCP, and DNS datasets, and 1000 ms for HTTP datasets, as suggested in \cite{indola_suitability_2023}.

The selected algorithms, Random Forest, Extreme Gradient Boosting (XGBoost), Light Gradient-Boosting Machine (LightGBM), and Categorical Boosting (CatBoost), are chosen for their effectiveness in handling structured data, robustness against noise, and ability to capture complex relationships within datasets. These models are widely used for predictive modeling due to their strong generalization capabilities and competitive performance across various domains\cite{Mahesh2019,Huang2019,Ray2019,Fernandez2019,Hancock2020,jung2022machine}. 

Random Forest, an ensemble learning method, aggregates predictions from multiple decision trees to enhance accuracy and mitigate overfitting \cite{Mahesh2019}. The model is configured with 100 estimators and a fixed random seed to ensure reproducibility. Its scalability and versatility make it suitable for both classification and regression tasks.

XGBoost is a powerful gradient-boosting algorithm, optimized for speed and performance, using an iterative approach to minimize errors and enhance predictions. Its regularization and efficient handling of sparse data make it highly effective on large datasets\cite{Ray2019}. The configuration employs 100 estimators, a fixed random seed, and the mean squared error (MSE) loss function for regression.

LightGBM is a gradient-boosting framework optimized for efficiency, especially on large datasets. Its ability to handle categorical data and scale well makes it highly effective for predictive tasks. It uses leaf-wise tree growth instead of level-wise splitting, improving both training speed and accuracy\cite{Huang2019}. In this study, the model is configured with 100 boosting estimators and a fixed random seed. 

CatBoost is a gradient-boosting library designed to efficiently handle categorical features with minimal preprocessing \cite{Hancock2020}. Its robustness on structured data makes it well-suited for tabular prediction tasks. The model is configured with 100 iterations, a learning rate of 0.1, and a tree depth of 6 to balance learning efficiency and generalization. Training output is suppressed using the \texttt{silent} parameter, and a fixed random seed is used to ensure reproducibility. 

The performance of ML models is evaluated, and the best model is selected based on Root Mean Squared Error (RMSE), R², Mean Absolute Error (MAE), precision, and recall. RMSE is a common regression metric providing a straightforward way to quantify the difference between predicted and actual values, with errors expressed in the same units as the target variable. R², the coefficient of determination, measures how well the predicted values match the actual data. R² ranges from $-\infty$ to 1, where 1 indicates perfect predictions. MAE evaluates the average absolute difference between the predicted and actual values \cite{jung2022machine,Plevris2022,Botchkarev2018}.

For classification tasks, precision and recall are essential metrics. Precision is the ratio of true positives to total predicted positives, while recall is the ratio of true positives to actual positives. Both metrics range from 0 to 1, with higher values indicating fewer false positives (precision) or false negatives (recall). Achieving a balance between precision and recall is crucial for tasks where both over-prediction and under-prediction incur significant costs\cite{Botchkarev2018}.

\section{Modeling Results and Dataset Validation}

The performance of four ML algorithms used to model delays in PR mode is presented in Table \ref{tab:all_delays_merged_column}. Because the primary objective is to model the observed delays under packet replication accurately, evaluation metrics are calculated by comparing predictions to the actual PR-mode measurements. Accordingly, emphasis is placed on the overall results rather than on separate training and testing metrics.

Random Forest performs best on the {\bf Position Report dataset}, achieving the lowest RMSE of 8.18, the highest R² of 0.84, and the lowest MAE of 3.22, as well as balanced Precision (0.44) and Recall (0.80) for extreme cases. For \textbf{MA request}, XGBoost outperforms the other models with the lowest RMSE of 8.73 and the highest R² of 0.97. However, its MAE (5.44) is higher than Random Forest’s; it maintains robust anomaly handling with a precision of 0.50 and a perfect recall of 1.00. 

For {\bf TCP dataset}, Random Forest provides the best overall metrics, including an RMSE of 8.23, R² of 0.57, and MAE of 2.57, alongside balanced extreme-case precision and recall (both 0.60). Also for {\bf HTTP}, Random Forest again exhibits the best error metrics (RMSE 71.57, R² 0.50, MAE 3.95), but struggles with outliers as reflected in its lower Precision (0.25), though it still achieves a Recall of 0.60. Finally, for {\bf DNS}, Random Forest attains the lowest RMSE of 29.09, highest R² of 0.35, and smallest MAE of 2.83, with a Precision of 0.60 and Recall of 0.75 for extreme cases. 

\begin{table}[htbp]
\centering
\caption{Performance metrics for all delays in PR Mode (Full dataset).}
\label{tab:all_delays_merged_column}
\resizebox{\columnwidth}{!}{%
\begin{tabular}{l l c c c c}
\toprule

\textbf{Delay} & \textbf{Metric} & \textbf{Random Forest} & \textbf{XGBoost} & \textbf{LightGBM} & \textbf{CatBoost} \\
\midrule

\multirow{5}{*}{\textbf{Position}} 
 & RMSE                     & 8.18  & 9.51  & 15.54 & 10.97 \\
 & R\textsuperscript{2}     & 0.84  & 0.78  & 0.42  & 0.71  \\
 & MAE                      & 3.22  & 5.88  & 6.43  & 6.57  \\
 & Precision      & 0.44  & 0.50  & 0.38  & 0.50  \\
 & Recall         & 0.80  & 0.60  & 0.60  & 0.60  \\
\midrule

\multirow{5}{*}{\textbf{MA}}
 & RMSE                     & 30.78 & 8.73  & 40.13 & 17.81 \\
 & R\textsuperscript{2}     & 0.65  & 0.97  & 0.40  & 0.88  \\
 & MAE                      & 4.20  & 5.44  & 8.10  & 7.80  \\
 & Precision      & 0.33  & 0.50  & 0.50  & 0.25  \\
 & Recall         & 1.00  & 1.00  & 1.00  & 1.00  \\
\midrule

\multirow{5}{*}{\textbf{TCP}}
 & RMSE                     & 8.23  & 10.45 & 10.85 & 11.01 \\
 & R\textsuperscript{2}     & 0.57  & 0.31  & 0.26  & 0.24  \\
 & MAE                      & 2.57  & 4.78  & 5.14  & 5.27  \\
 & Precision      & 0.60  & 0.43  & 0.50  & 0.43  \\
 & Recall         & 0.60  & 0.60  & 0.40  & 0.60  \\
\midrule

\multirow{5}{*}{\textbf{HTTP}}
 & RMSE                     & 71.57 & 90.76 & 90.82 & 90.79 \\
 & R\textsuperscript{2}     & 0.50  & 0.20  & 0.20  & 0.20  \\
 & MAE                      & 3.95  & 7.82  & 9.13  & 9.50  \\
 & Precision     & 0.25  & 0.21  & 0.21  & 0.21  \\
 & Recall         & 0.60  & 0.60  & 0.60  & 0.60  \\
\midrule

\multirow{5}{*}{\textbf{DNS}}
 & RMSE                     & 29.09 & 29.77 & 32.97 & 30.26 \\
 & R\textsuperscript{2}     & 0.35  & 0.32  & 0.17  & 0.30  \\
 & MAE                      & 2.83  & 5.16  & 5.75  & 5.79  \\
 & Precision     & 0.60  & 0.75  & 0.50  & 0.67  \\
 & Recall         & 0.75  & 0.75  & 0.50  & 0.50  \\

\bottomrule
\end{tabular}
}
\end{table}

Comparing actual data measured in PR Mode (2000 km) with estimates from ML algorithms revealed a consistent pattern. In nearly all delay types, critical points (defined as exceeding predefined {\em critical} thresholds (500ms or 1000ms)) were detected by the ML models slightly earlier than their actual occurrence and often persisted for several seconds longer. This indicates that critical conditions are identified over a broader time window. While real-world effects are observed for a few seconds, the ML model classifies the entire duration as critical. This suggests that the model does not fully capture all real-world conditions, resulting in a broader classification of the critical period rather than pinpointing the exact moment the threshold is exceeded. Although this may slightly affect specific evaluation metrics, it is not considered a significant issue, as the objective is to capture the worst-case statistical performance across the railway network. Additionally, train speed was identified as one of the most influential features in accurately estimating PR-mode delays.

\begin{figure}[htbp]
    \centering
    \includegraphics[width=0.40\textwidth]{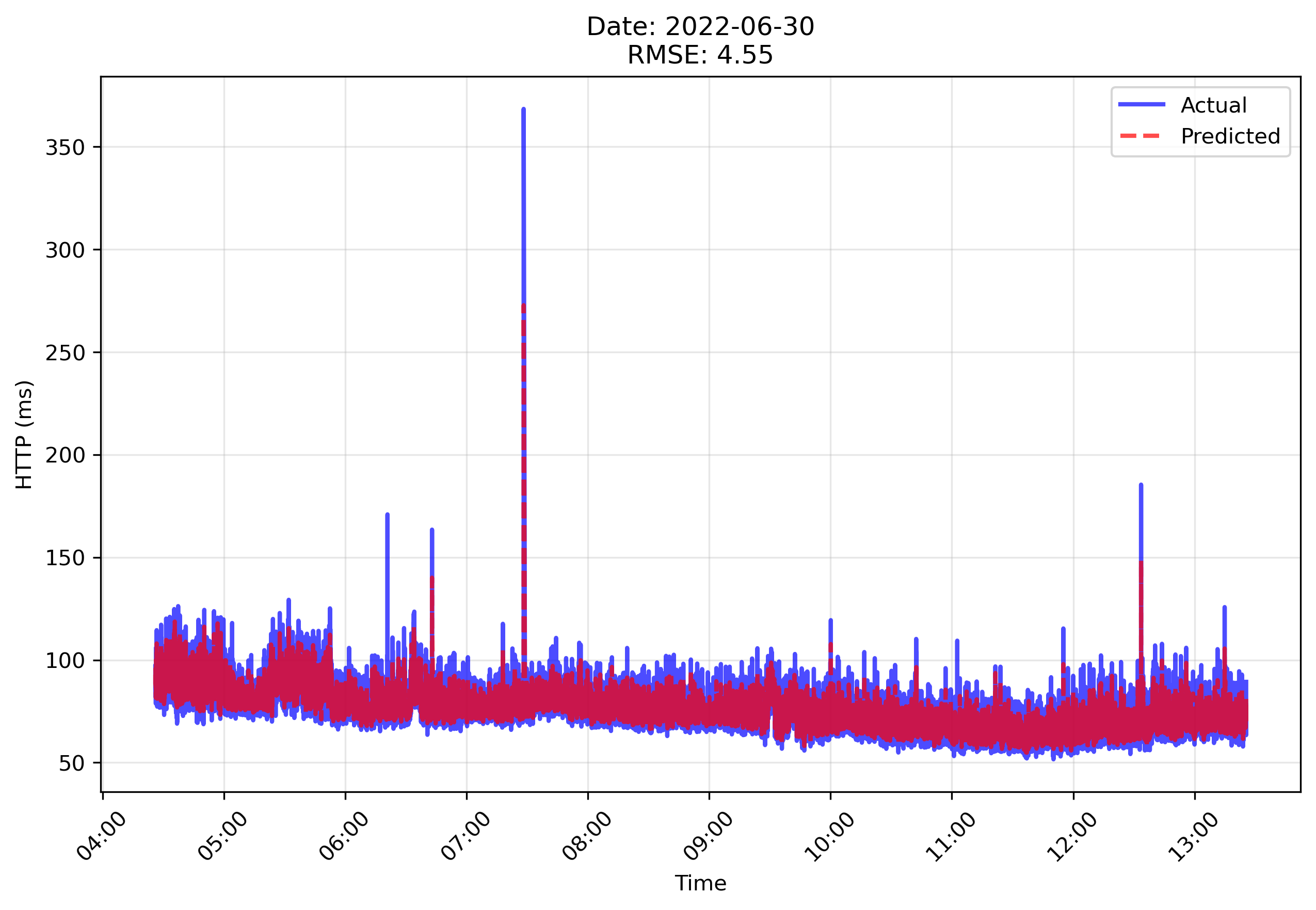}
    \caption{Actual vs. predicted HTTP delay on June 30, 2022}
    \label{fig:Actual vs. predicted HTTP}
\end{figure}

\begin{figure}[htbp]
    \centering
    \includegraphics[width=0.40\textwidth]{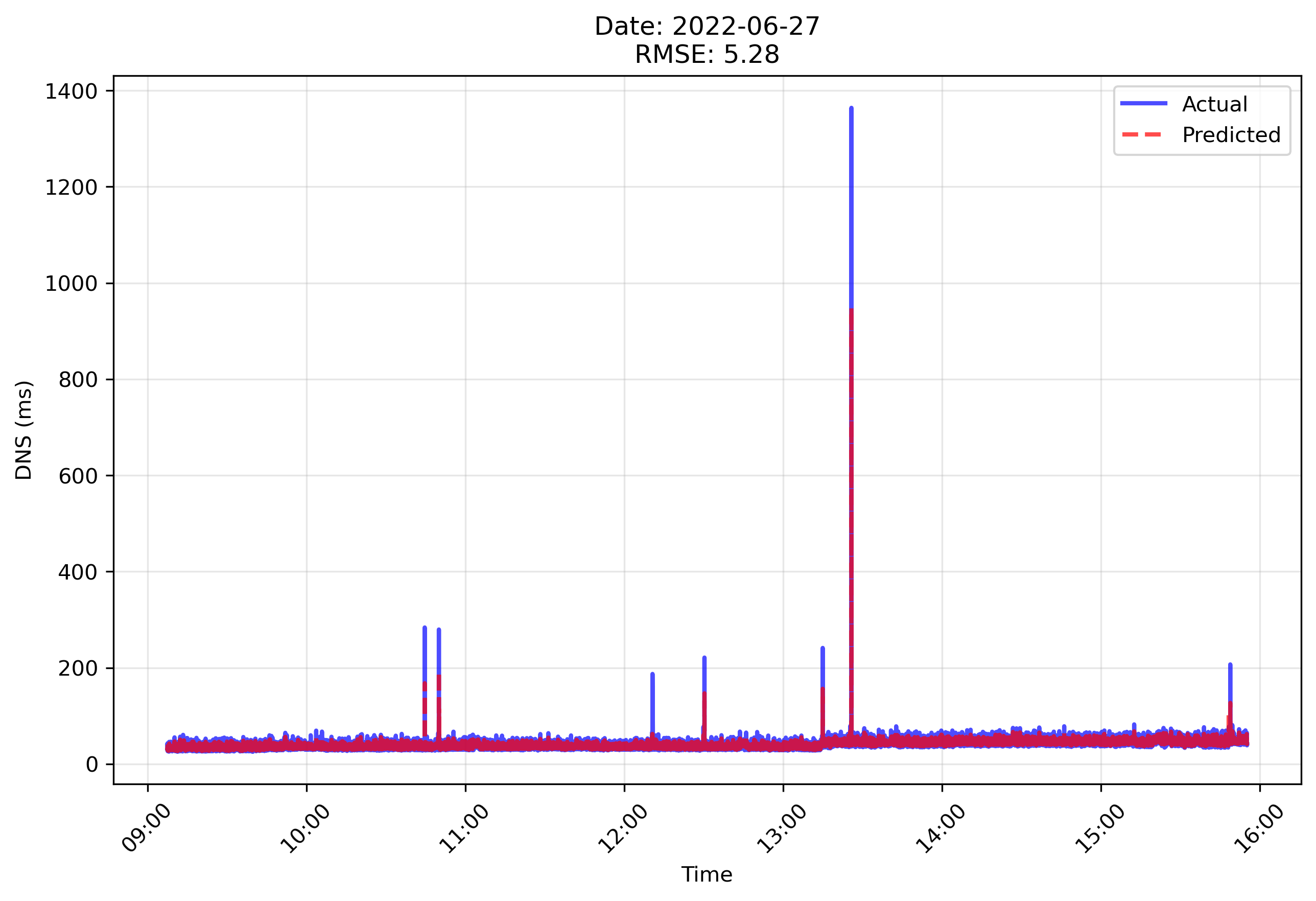}
     \caption{Actual vs. predicted DNS delay on June 27, 2022}
    \label{fig:Actual vs. predicted DNS}
\end{figure}

Figures \ref{fig:Actual vs. predicted HTTP} and \ref{fig:Actual vs. predicted DNS} show daily HTTP and DNS delay plots on two different days. Blue lines represent actual PR mode measurements, while red dashed lines indicate predictions from the best-performing ML model (Random Forest). The model effectively captures the main trends and timing of significant delay spikes in the PR mode data. Each plot also includes the RMSE value for the corresponding day, providing a quantitative measure of prediction accuracy.

\begin{figure}[htbp]
    \centering
    \includegraphics[width=0.5\textwidth]{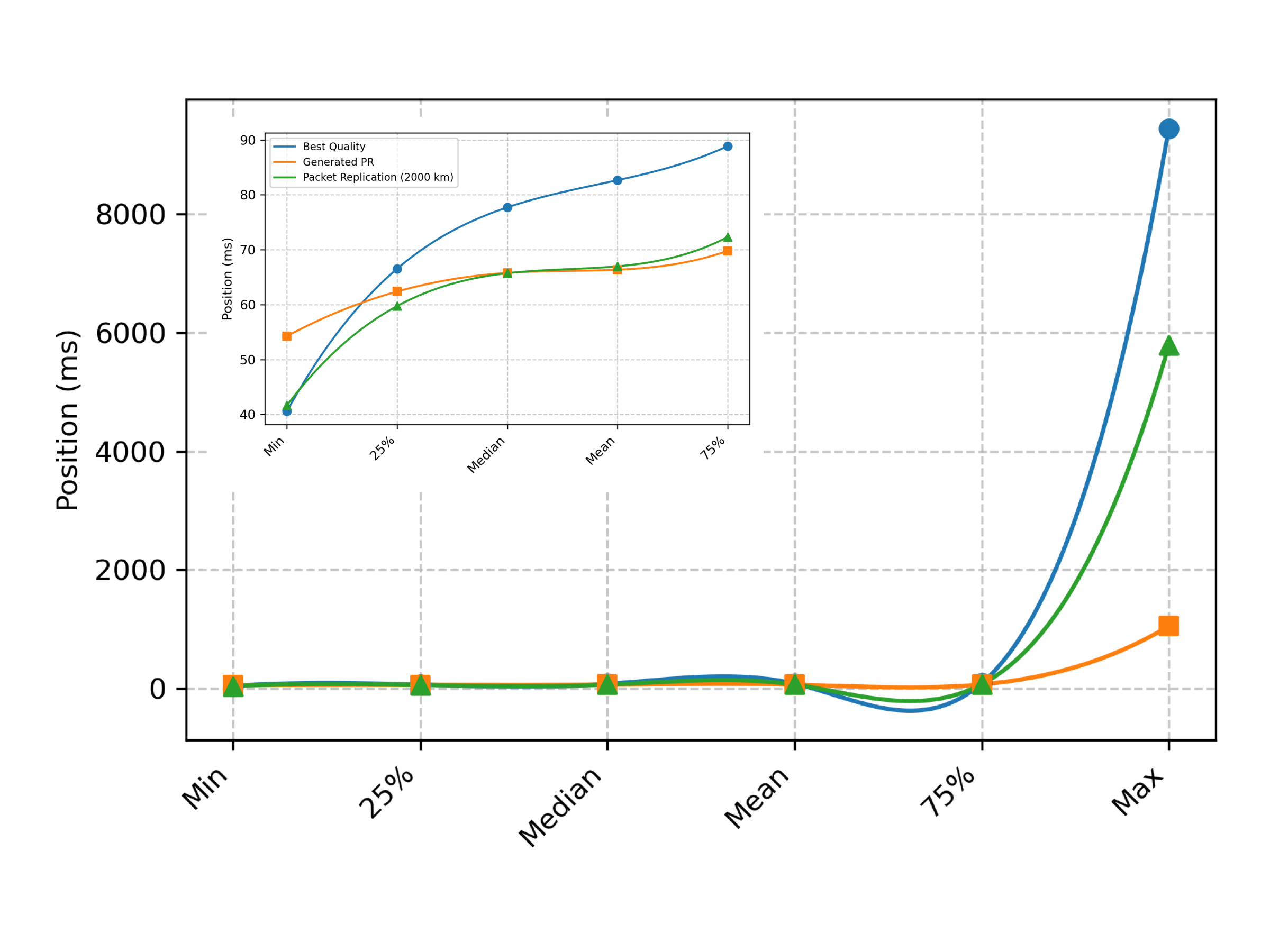}
    \caption{Statistical summary of Position Report across different scenarios.}
    \label{fig:position_delay}
\end{figure}

\begin{figure}[htbp]
    \centering
    \includegraphics[width=0.5\textwidth]{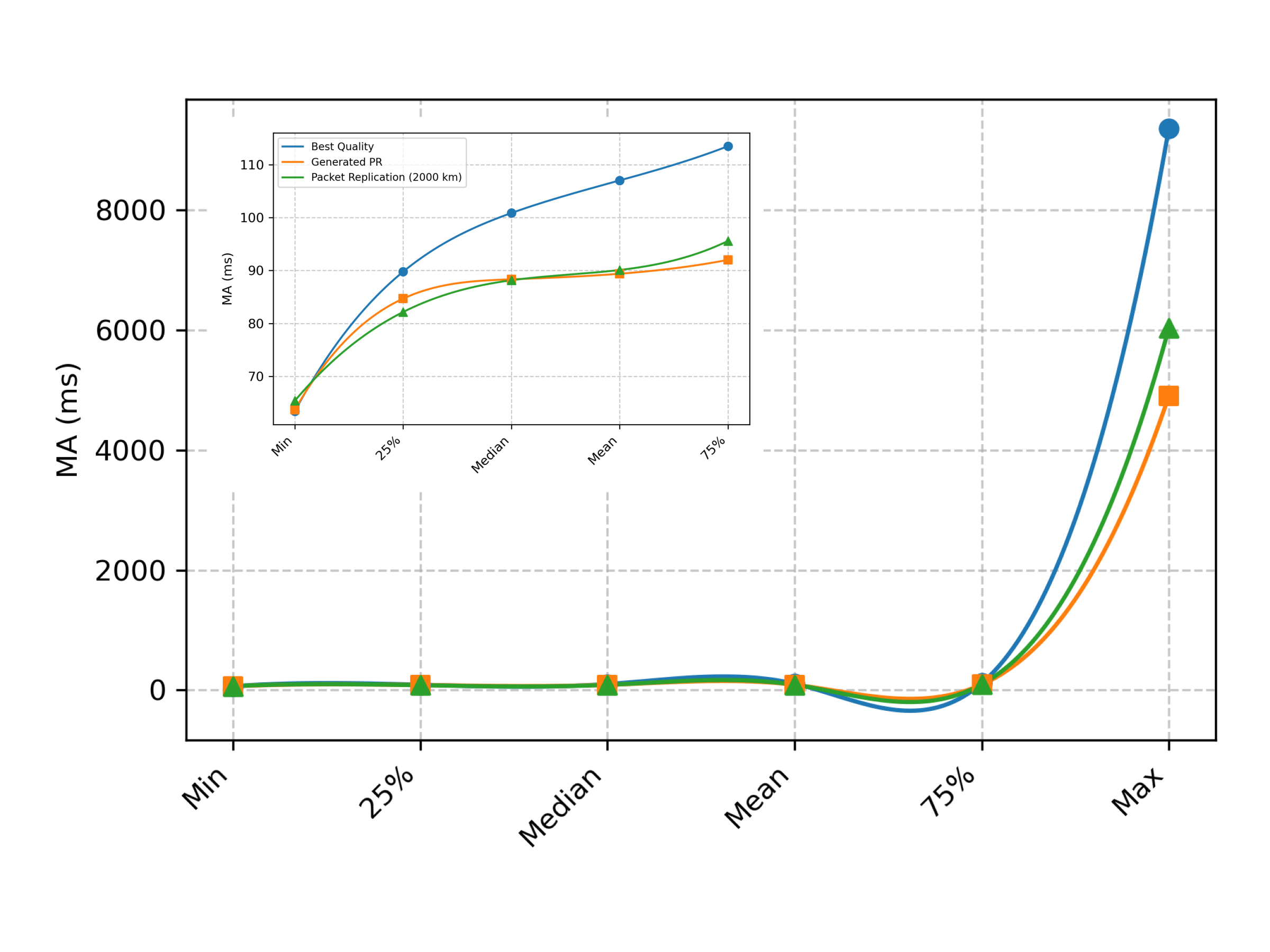}
    \caption{Statistical summary of MA Request across different scenarios.}
    \label{fig:ma_delay}
\end{figure}

The best-performing machine learning algorithm for each delay type is used to generate the corresponding PR-mode dataset for segments initially measured in BQ mode. Since no measured PR-mode data exists for these segments, standard evaluation metrics such as RMSE and MAE cannot be applied. Instead, the generated datasets are evaluated by comparing their statistical characteristics to actual PR-mode measurements. Due to space constraints, the focus is placed on Position Report and MA Request, as shown in Figures \ref{fig:position_delay}–\ref{fig:ma_delay}. TCP, HTTP, and DNS results exhibit similar trends and are therefore omitted for conciseness.

Each delay type is evaluated under three scenarios: BQ mode, ML-generated PR mode, and measured PR mode (based on 2000 km of railway data). Blue circles, orange squares, and green triangles represent BQ, generated PR, and measured PR data, respectively.

Each plot includes key statistical indicators: minimum, 25th percentile, median, mean, 75th percentile, and maximum. A magnified inset is provided to aid in comparison in the lower delay ranges. The results show strong alignment between the ML-generated and measured PR-mode data regarding central tendencies, indicating that the models effectively replicate real-world delay behavior. Some differences appear at the extremes, with measured data showing higher maximum delays and greater variability, while ML-generated data is more tightly clustered. In contrast, BQ mode consistently yields higher delays across all metrics, highlighting its distinct performance profile compared to the PR-based scenarios.

\section{Evaluating Mobile Network Reliability}

The reliability of the Finnish public LTE network for railway communication is evaluated by analyzing the percentage of critical delay events across BQ, PR, and ML-generated PR modes. The results are presented in Table \ref{tab:delays_above_500_combined}.

The analysis of data points exceeding the predefined thresholds \cite{indola_suitability_2023} across five delay types shows that such high-delay events are significantly more common in BQ mode. This is likely due to the artificial delays introduced in BQ mode, which do not accurately represent real-world network conditions. In contrast, the Measured PR mode dataset contains very few critical delays. Furthermore, the generated datasets in PR mode using the best-performing ML models closely mirror the patterns observed in the actual PR mode measurements. This indicates that the models have effectively learned and replicated PR-mode delay behavior.

The findings confirm that the public mobile network is capable of supporting ETCS-related communication with high reliability, meeting the strict delay requirements in over 99.9\% of cases. This strongly supports the feasibility of using public LTE infrastructure, when enhanced with packet replication, for railway operations. 

\begin{table}[htbp]
\centering
\caption{Data points exceeding 500 ms across delay types and scenarios.}
\label{tab:delays_above_500_combined}
\resizebox{\columnwidth}{!}{%
\begin{tabular}{l l c c}
\toprule
\textbf{Delay} & \textbf{Measurement Mode} & \textbf{Count} & \textbf{Percentage in Whole Data} \\
\midrule

\multirow{3}{*}{\textbf{Position Report}} 
 & Best Quality                 & 1409 & 0.19\% \\
 & Generated PR Mode            & 21   & 0.003\% \\
 & Measured PR Mode (2000 km)   & 5    & 0.004\% \\
\midrule

\multirow{3}{*}{\textbf{MA}} 
 & Best Quality                 & 166  & 0.21\% \\
 & Generated PR Mode            & 42   & 0.05\% \\
 & Measured PR Mode (2000 km)   & 1    & 0.007\% \\
\midrule

\multirow{3}{*}{\textbf{TCP}} 
 & Best Quality (4000 km)                 & 1175 & 0.16\% \\
 & Generated PR Mode (4000 km)            & 0    & 0.00\% \\
 & Measured PR Mode (2000 km)             & 5    & 0.004\% \\
\midrule

\multirow{3}{*}{\textbf{HTTP}} 
 & Best Quality                 & 983  & 0.13\% \\
 & Generated PR Mode            & 668  & 0.09\% \\
 & Measured PR Mode (2000 km)   & 5    & 0.004\% \\
\midrule

\multirow{3}{*}{\textbf{DNS}} 
 & Best Quality                 & 1469 & 0.19\% \\
 & Generated PR Mode            & 46   & 0.01\% \\
 & Measured PR Mode (2000 km)   & 4    & 0.003\% \\

\bottomrule
\end{tabular}
}
\end{table}

\begin{figure}[htbp]
    \centering
    \includegraphics[width=0.35\textwidth]{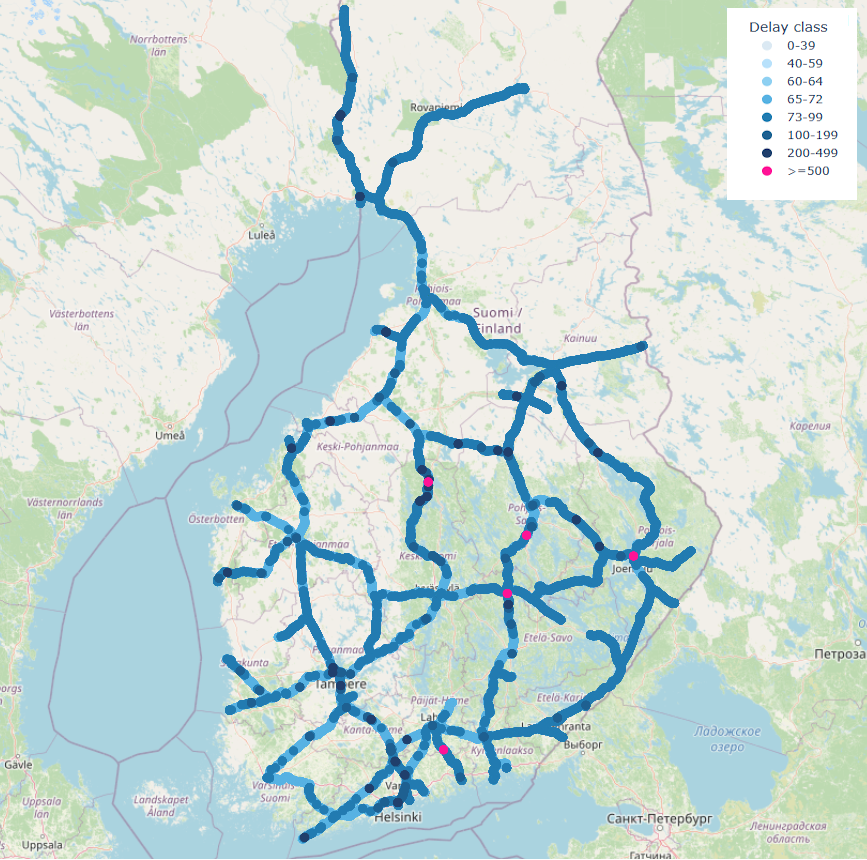}
    \caption{Generated Position Report in Packet Replication mode.}
    \label{fig:position_Ge_BQ}
\end{figure}

 As an example, Figures \ref{fig:position_BQ} and \ref{fig:position_Ge_BQ} illustrate the delay results for the Position Report in BQ mode and ML-generated PR mode, respectively. Data points are color-coded from light to dark blue for delays up to 500 ms, while pink indicates critical delays exceeding 500 ms. The number of critical delay events in the generated PR-mode dataset is significantly lower than in the measured BQ mode, aligning with the results in Table \ref{tab:delays_above_500_combined}. This demonstrates that the developed ML model can effectively transform BQ-mode measurements into PR-mode statistics with high \textit{statistical} accuracy.

\begin{figure}[htbp]
    \centering
    \includegraphics[width=0.35\textwidth]{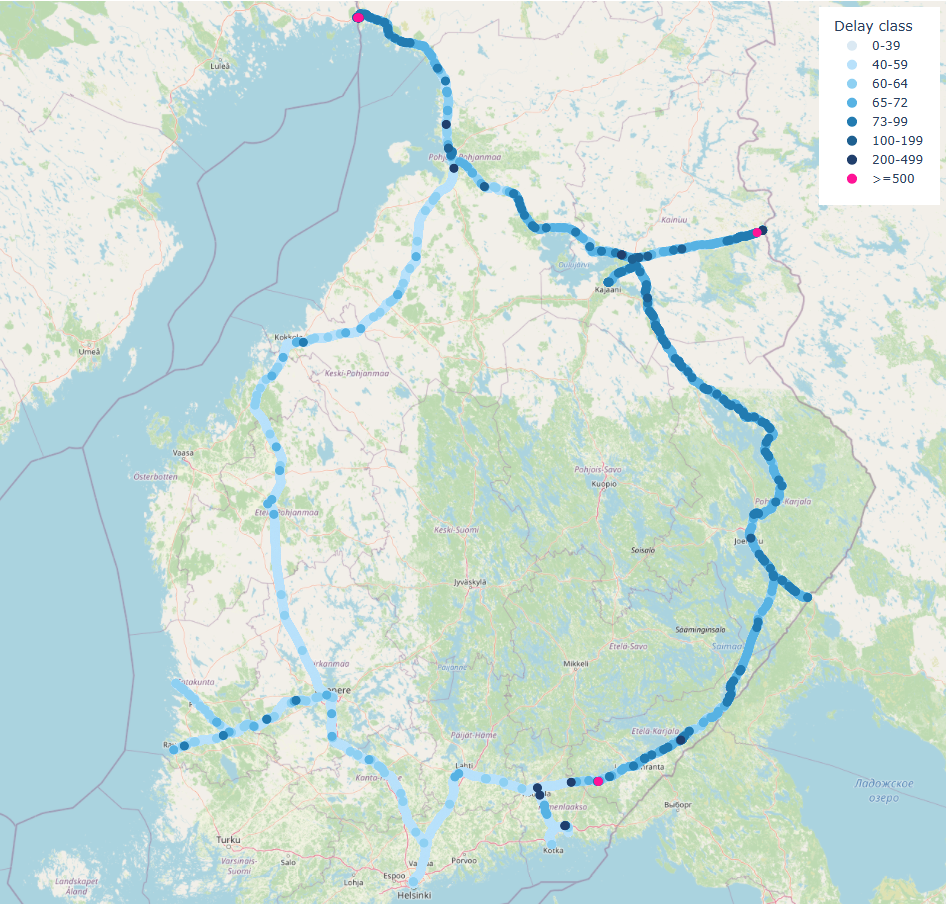}
    \caption{DNS (Actual Packet Replication mode).}
    \label{fig:DNS_PR}
\end{figure}

\textcolor{black}{Data in PR mode was collected along a 2,000 km railway border ring in Finland. Figure \ref{fig:DNS_PR} displays the railway segments measured in PR mode, where data points represent DNS delays. Delays up to 500 ms are shown in light to dark blue, while pink indicates critical delays exceeding 500 ms. The results in Table \ref{tab:Regional_test_statistics} highlight notable regional differences in network delay performance between the eastern and western sections of the ring. Across all measured metrics, the eastern region consistently exhibits higher mean delays. For instance, HTTP delays average 87.60 ms in the east compared to 74.20 ms in the west. Similarly, Position Report delays are 70.39 ms versus 63.57 ms, and DNS delays are 45.82 ms versus 39.85 ms, respectively. These findings reveal a persistent regional performance gap, with higher delays consistently observed in the eastern part of the network.}

\textcolor{black}{The modeled PR-mode delays, presented in Table \ref{tab:predicted_Regional_test_statistics}, reflect a similar pattern. The ML-generated metrics closely align with the measured values and successfully capture the east-west delay disparity. This consistency further validates the accuracy of the trained models and confirms the presence of a regional bias in QoS.}

\begin{table}[htbp]
\centering
\caption{Regional statistics for actual delays.}
\label{tab:Regional_test_statistics}
\resizebox{\columnwidth}{!}{%
\begin{tabular}{l l c c c c c c}
\toprule
\textbf{Test Name} & \textbf{Region} & \textbf{Mean} & \textbf{Min} & \textbf{25\%} & \textbf{50\%} & \textbf{75\%} & \textbf{Max} \\
\midrule

\multirow{2}{*}{\textbf{Position Report (ms)}} 
 & East & 70.39 & 43.58 & 64.28 & 69.28 & 75.70 & 2,130.14 \\
 & West & 63.57 & 41.67 & 57.41 & 61.73 & 67.62 & 5,787.29 \\
\midrule

\multirow{2}{*}{\textbf{MA (ms)}} 
 & East & 93.71 & 68.90 & 85.87 & 91.63 & 99.00 & 6,029.01 \\
 & West & 86.47 & 65.37 & 79.86 & 84.77 & 91.20 & 492.43 \\
\midrule

\multirow{2}{*}{\textbf{TCP (ms)}} 
 & East & 40.19 & 23.36 & 35.38 & 39.11 & 44.08 & 1,901.13 \\
 & West & 33.02 & 21.08 & 28.06 & 31.66 & 36.71 & 1,645.39 \\
\midrule

\multirow{2}{*}{\textbf{HTTP (ms)}} 
 & East & 87.60 & 52.97 & 79.33 & 86.88 & 94.71 & 31,569.19 \\
 & West & 74.20 & 47.89 & 65.80 & 72.18 & 80.04 & 8,939.65 \\
\midrule

\multirow{2}{*}{\textbf{DNS (ms)}} 
 & East & 45.82 & 27.07 & 40.71 & 44.72 & 49.88 & 10,071.37 \\
 & West & 39.85 & 25.34 & 34.18 & 38.42 & 43.65 & 6,249.93 \\
\bottomrule
\end{tabular}
}
\end{table}

\begin{table}[htbp]
\centering
\caption{Regional statistics for modeled delays.}
\label{tab:predicted_Regional_test_statistics}
\resizebox{\columnwidth}{!}{%
\begin{tabular}{l l c c c c c c}
\toprule
\textbf{Test Name} & \textbf{Region} & \textbf{Mean} & \textbf{Min} & \textbf{25\%} & \textbf{50\%} & \textbf{75\%} & \textbf{Max} \\
\midrule

\multirow{2}{*}{\textbf{Position Report (ms)}} 
 & East & 70.42 & 47.24 & 66.06 & 70.32 & 74.89 & 1,377.42 \\
 & West & 63.73 & 46.49 & 58.79 & 62.30 & 67.09 & 3,995.78 \\
\midrule

\multirow{2}{*}{\textbf{MA (ms)}} 
 & East & 93.00 & 72.02 & 89.17 & 92.33 & 95.19 & 6,005.04 \\
 & West & 87.24 & 70.35 & 83.30 & 86.38 & 90.12 & 666.70 \\
\midrule

\multirow{2}{*}{\textbf{TCP (ms)}} 
 & East & 40.16 & 23.97 & 36.74 & 39.92 & 43.51 & 256.66 \\
 & West & 33.09 & 22.41 & 29.30 & 32.14 & 35.83 & 1,021.55 \\
\midrule

\multirow{2}{*}{\textbf{HTTP (ms)}} 
 & East & 87.85 & 56.46 & 81.21 & 87.77 & 93.84 & 18,042.13 \\
 & West & 74.34 & 52.92 & 67.11 & 72.17 & 79.08 & 5,754.01 \\
\midrule

\multirow{2}{*}{\textbf{DNS (ms)}} 
 & East & 45.66 & 28.95 & 42.08 & 45.52 & 49.29 & 947.92 \\
 & West & 39.92 & 26.80 & 35.47 & 38.72 & 43.01 & 3,785.61 \\
\bottomrule
\end{tabular}
}
\end{table}

\textcolor{black}{This regional disparity is also visually evident in Figure \ref{fig:DNS_PR}, where delays are generally higher in the eastern segment, reinforcing the statistical findings. Although delays remain within acceptable limits, this discrepancy may require attention prior to operational deployment.}

\section{Conclusion}

This study evaluates the performance of four ML algorithms in modeling delays across various network conditions, particularly focusing on the ETCS communication in PR mode. The findings reveal that the ML models effectively capture delay behavior, with Random Forest performing best in Position Report delay prediction and XGBoost excelling in MA Request delay prediction. The models consistently identified critical delay points, albeit with a broader time window, suggesting that the conditions leading to these delays exist over extended periods.

It is hypothesized that handovers play a crucial role in delay spikes, based on the observation that train velocity provides stronger explanatory power in the ML models than the train’s absolute location. Further investigation into this aspect is warranted to validate the hypothesis and refine the understanding of delay-inducing factors.

The results indicate a high statistical similarity between the measured data and the data generated by the selected ML models. This suggests that the models effectively capture the underlying patterns of delays in PR Mode. Moreover, the study confirms that public mobile networks in Finland are capable of reliably supporting ETCS traffic in over 99.9\% of locations. 

The developed ML modeling approach can enhance the efficiency of measurement campaigns by generating digital twin-type delay data statistics from BQ measurements. Future work will integrate the ML model with digitally simulated data from the railway network to test its ability to reliably generate delay information from simulated ray-traced and dynamic railway data.

\section*{Acknowledgment}
This study was conducted as part of the Digirail project, using data provided within its framework.

\vspace{12pt}
\end{document}